\documentclass[preprint,showpacs,preprintnumbers]{revtex4}
\usepackage[active]{srcltx} 
\usepackage{amsmath}
\usepackage{graphicx}

\newcommand{\be}{\begin{equation}}
\newcommand{\ee}{\end{equation}}
\def\Jb{{\mathbf J}}

\begin{document}

\title{Statistical Analysis of Crossed Undulator for Polarization Control in a SASE FEL}

\author{Yuantao Ding and Zhirong Huang}

\affiliation{Stanford Linear Accelerator Center, Menlo Park, CA
94025}


\pacs{41.60.Cr}


\begin{abstract}

There is a growing interest in producing intense, coherent x-ray
radiation with an adjustable and arbitrary polarization state. In
this paper, we study the crossed undulator scheme (K.-J. Kim, Nucl.
Instrum. Methods A {\bf 445}, 329 (2000)) for rapid polarization
control in a self-amplified spontaneous emission (SASE) free
electron laser (FEL). Because a SASE source is a temporally chaotic
light, we perform a statistical analysis on the state of
polarization using FEL theory and simulations. We show that by
adding  a small phase shifter and a short (about 1.3 times the FEL
power gain length), $90^\circ$ rotated planar undulator after the
main SASE planar undulator, one can obtain circularly polarized
light -- with  over 80\% polarization -- near the FEL saturation.

\end{abstract}

\maketitle

\section{Introduction}

Several x-ray free electron lasers (FELs) based on self-amplified
spontaneous emission (SASE) are being developed worldwide as
next-generation light sources~\cite{LCLS,TESLA,SCSS}. In the soft
x-ray wavelength region, polarization control (from linear to
circular) is highly desirable in studying ultrafast magentic
phenomena and material science. The x-ray FEL is normally linearly
polarized based on planar undulators. Variable polarization could in
principle be provided by employing an APPLE-type
undulator~\cite{apple}. However, its mechanical tolerance for lasing
at x-ray wavelengths has not been demonstrated, and its focusing
property may change significantly when its polarization is altered.
An alternative approach for polarization control is the so-called
``crossed undulator" (or ``crossed-planar undulator"), which is the
subject of this paper.

The crossed-planar undulator was proposed by K.-J Kim to generate
arbitrarily polarized light in synchrotron
radiation~\cite{kim_nim219} and FEL sources~\cite{kim_nim445}. It is
based on the interference of horizontal and vertical radiation
fields generated by two adjacent planar undulators in a crossed
configuration (see Fig.~\ref{layout}). A phase shifter between the
undulators is used to delay the electron beam and hence to control
the final polarization state. For incoherent radiation sources, the
radiation pulses generated in two adjacent undulators by each
electron do not overlap in time. Thus, a monochromator after the
second undulator is required to stretch both pulses temporally in
order to achieve interference. The degree of polarization is limited
by beam emittance, energy spread, and the finite resolution of the
monochromator, as studied in a series of experiments at
BESSY~\cite{bessy1,bessy2}. On the other hand, for completely
coherent radiation sources (such as generated from a seeded FEL
amplifier or an FEL oscillator), the interference occurs due to the
overlap of two radiation components in the second
undulator~\cite{kim_nim445}. A recent crossed-undulator experiment
at the Duke storage ring FEL reported controllable polarization
switches with a nearly 100\% total degree of
polarization~\cite{duke}.

It is well-known in the FEL community that SASE light is
transversely coherent but temporally chaotic due to the shot noise
startup. Thus, the effectiveness of the crossed undulator for
polarization control deserves a detailed study. In this paper,
starting with one-dimensional (1D) FEL theory, we calculate both
radiation components and generalize the results of
Ref.~\cite{kim_nim445} to the case of SASE. We then determine the
required length of the second undulator in order to produce the same
average power as that produced in the first undulator. We show that
the degree of polarization can be determined by the time correlation
of the two radiation fields and compute its asymptotic expression in
the high-gain limit. The analytical results are compared with 1D
SASE simulations after a proper statistical averaging. Finally,
three-dimensional (3D) effects and simulation results are also
discussed.

\section{Field calculation}

\begin{figure}[tb]
\centering
\includegraphics[scale=0.9]{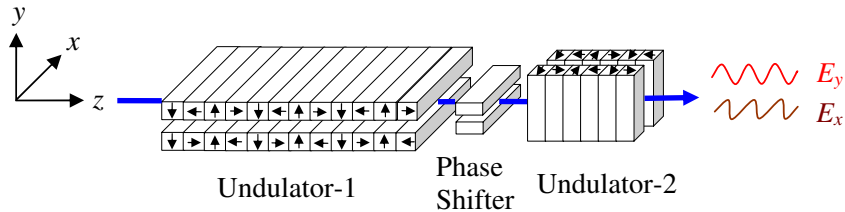}
\caption{(color) Schematic of the crossed undulator for polarization
control} \label{layout}
\end{figure}

Figure~\ref{layout} shows a schematic of the crossed undulator
applied to a SASE FEL. In the first planar undulator with a total
length $L_1$, spontaneous radiation is amplified to generate
horizontally polarized SASE field $E_{x}$. In the second undulator
(of length $L_2$) that is rotated 90$^\circ$ with respect to the
first one, $E_x$ propagates freely without interacting with the
electron beam, while a vertically polarized radiation field $E_{y}$
is produced by the micro-bunched beam. A simple phase shifter such
as a four-dipole chicane placing between the two undulators can
slightly delay the electrons in order to adjust the relative phase
of the two polarization components.

In this section, we determine both SASE field components generated
by the crossed undulator. Let $E(z,t)$ be the complex but slowly
varying electric field at undulator distance $z$ and time $t$. We
write
    \be
    E(z,t)=\int {\omega_1 d\nu\over \sqrt{2\pi}} E_\nu (z) e^{i\Delta \nu \left[(k_1+k_u)
    z-\omega_1 t\right]}\,, \label{eq:fourier}
    \ee
where $\omega_1=k_1 c$ is the fundamental resonant frequency
corresponding to the average beam energy ($c$ is the speed of
light); $\nu=\omega/\omega_1$ and $\Delta \nu=\nu-1$ is the relative
frequency detuning, $k_u=2\pi/\lambda_u$ with $\lambda_u$ the
undulator period. Following Refs.~\cite{Kim_nim86,wang}, the 1D FEL
interaction starting from shot noise can be described by the coupled
Maxwell-Klimontovich equations. In the small signal regime before
FEL saturation, the equations can be linearized and solved by the
Laplace transformation:

    \begin{align}\label{basic solution}
     E_{\nu}(z)&=\oint\frac{d\mu}{2\pi i}(-i2\rho k_u)e^{-i2\rho\mu
    k_{u}z}E_{\nu,\mu}\,,\nonumber
    \\ F_{\nu}(z)&=\oint\frac{d\mu}{2\pi i}e^{-i2\rho\mu
    k_{u}z}\frac{\kappa_{1}E_{\nu,\mu}dV/d\eta-F_{\nu}(0)}
    {(\eta/\rho-\mu )}\,,
    \end{align}
where
    \begin{align}\label{dispersion} E_{\nu,\mu}&=\frac{i}{2\rho
    k_uD(\mu)}\left(E_{\nu}(0)+\frac{i\kappa_{2}n_{0}}{2\rho k_u}\int
    d\eta\frac{F_\nu(0)}{\eta/\rho-\mu}\right)\,,\nonumber\\
    D(\mu)&=\mu-\frac{\Delta\nu}{2\rho}-\int d\eta
    \frac{V(\eta)}{(\eta/\rho-\mu)^2}\,.
    \end{align}
Here $E_\nu$ and $F_\nu$ are, respectively, the Fourier components
of the electric field and of the Klimontovich distribution function
that describes the discrete electrons in longitudinal phase space,
with $E_{\nu}(0)$ and $F_{\nu}(0)$ the Fourier components of the
initial conditions; $D(\mu)=0$ determines the FEL dispersion
relation where $\mu$ is the Laplace parameter. In addition,
parameter $\rho$ is the dimensionless FEL Pierce
parameter~\cite{BPN}, $V(\eta)$ is the electron energy distribution
with $\eta$ the relative energy deviation, $n_0$ is the electron
volume density; $\kappa_1=eK\textrm{[JJ]}/(4\gamma_0^2mc^2)$,
$\kappa_2=eK\textrm{[JJ]}/(2\epsilon_0\gamma_0)$, where $K$ is the
dimensionless undulator strength parameter, the Bessel function
factor $\textrm{[JJ]}$ is equal to $[J_0(\xi)-J_1(\xi)]$ with
$\xi=K^2/(4+2 K^2)$, $\gamma_0$ is the initial electron energy in
units of $mc^2$, and $\epsilon_0$ is the vacuum permittivity. Note
that the contour integration of $\mu$ in Eq.~(\ref{basic solution})
must enclose all singularities in the complex $\mu$ plane. Based on
this solution, we can calculate radiation field components in the
crossed undulator according to their initial conditions.

\subsection{Horizontal radiation field}

The radiation field $E_x$ in the first undulator develops from
electron shot noise, with the initial conditions\be
    E_\nu^x(0)=0\,, \quad \int F_\nu^x(0)d\eta=\frac{1}{N_\lambda}\sum^{N_e}_{j=1}e^{i\nu\omega_1 t_j(0)}\,,
\ee where $N_\lambda$ is the number of electrons in one radiation
wavelength, and $t_j(0)$ is the random arrival time of the $j^{th}$
electron at the entrance to the first undulator. We assume the first
undulator operates in the exponential growth regime. In this regime,
the dispersion relation has a solution $\mu_0$ with a positive
imaginary part that gives rise to an exponentially growing field
amplitude. For a cold beam with vanishing energy spread, we take
$V(\eta)=\delta(\eta)$ in Eq.~(\ref{basic solution}) and obtain
    \begin{align}\label{Exf}
    E_{\nu}^x(z)&=\frac{-i\kappa_2n_0}{2\rho k_uN_\lambda 3\mu_0}e^{-i\mu_0 2\rho
    k_uz}\sum^{N_e}_{j=1}e^{i\nu\omega_1 t_j} \quad \text{for $z\le L_1$.}
    \end{align}

In this high-gain regime, the electron distribution from
Eq.~(\ref{basic solution}) can be simplified as~\cite{kim_nim445}:
    \begin{align}\label{Fx}
    F_\nu^x (z) =
    \frac{i\kappa_1E_\nu^x(z)dV/d\eta}{2k_u(-\mu_0\rho+\eta)} \quad
    \text{for $z\le L_1$.}
    \end{align}
This electron distribution function will be used as an initial
condition for the calculation of the vertical radiation field as
follows.

\subsection{Vertical radiation field}

The radiation field $E_y$ in the second undulator is generated by
the pre-bunched electron beam in the first undulator. To control the
radiation polarization, the required path length delay of the phase
shifter chicane is on the order of the FEL wavelength. Such a weak
chicane does not have dispersive effects that could result in
micro-bunching, such as can be found for example in an optical
klystron (see, e.g., Ref.~\cite{OK}). Hence, the initial conditions
at the entrance of the second undulator is
    \be
    E_\nu^y(0)=0\,, \quad F_\nu^y(0)=F_\nu^x(L_1)\,.
    \ee

As the electron beam develops micro-bunching during the FEL
interaction in the first undulator, it will radiate coherently in
the second undulator. From discussions in Ref.~\cite{kim_nim445} and
simulation results shown in Sec.~\ref{sec:simu} below, the intensity
of $E_y$ can increase to the same level as that of $E_x$ in about
one gain length. Thus, for a relatively short second undulator, we
consider only coherent radiation and ignore any feedback of the
radiation on the electron beam. With this approximation, the third
term at the right hand side of $D(\mu)$ in Eq.~(\ref{dispersion})
can be dropped, and Eq.~(\ref{basic solution}) can now be written as
    \begin{align}\label{Enuy1}
    E_\nu^y(z_2)=&e^{i\phi}\oint\frac{d\mu}{2\pi i}\frac{e^{-i2\rho\mu
    k_{u}z_2}}{\mu-\Delta\nu/2\rho}\left[\frac{i\kappa_{2}n_{0}}{2\rho
    k_u}\int d\eta\frac{F_\nu^x(L_1)}{\eta/\rho-\mu}\right]
    \nonumber \\ =&-e^{i\phi}\oint\frac{d\mu}{2\pi i}\frac{e^{-i2\rho\mu
    k_{u}z_2}}{\mu-\Delta\nu/2\rho}\frac{
    E_\nu^x(L_1)}{\mu_0^2}\frac{\mu+\mu_0}{\mu^2}\,.
    \end{align}
Here $z_2$ is the undulator distance from the beginning of the
second undulator. The extra phase factor $e^{i\phi}$ is introduced
by the phase shifter just before the second undulator. In the last
step of Eq.~(\ref{Enuy1}), we have taken a cold beam with vanishing
energy spread and made use of the relation $\kappa_1\kappa_2 n_0
=4k_u^2 \rho^3$. Note that $\mu_0$ is the exponential growth
solution that satisfies $D(\mu_0)=0$ and is a function of the
detuning parameter $\Delta \nu$, i.e.,
    \be \label{eq:mu0}
    \mu_0\approx -{1\over 2} \left[1-{\Delta\nu\over 3\rho} + {(\Delta\nu)^2\over 36\rho^2}\right]
    +i {\sqrt{3}\over 2}\left[1- {(\Delta\nu)^2\over
    36\rho^2}\right]\,.
    \ee

Eq.~(\ref{Enuy1}) can be solved by the residue theorem:
    \begin{align}\label{Enuy2}
    E_\nu^y(z_2)=E_\nu^x(L_1)e^{i(\phi-\psi/2)}\textrm{sinc}\left({\psi\over
    2}\right)\frac{2i}{\mu_0^2}\left[\rho k_u z_2-\mu_0 e^{i\alpha}(\rho k_u z_2)^2
    \right]\,,
    \end{align}
where $\textrm{sinc}(x)=\sin (x)/x$, $\psi=\Delta\nu k_uz_2$, and
    \be
    \alpha=\arctan\left[\frac{\textrm{sin}(\psi/2)}{\textrm{sinc}(\psi/2)-\textrm{cos}(\psi/2)}\right]\,.
    \ee
Note that $\alpha=\pi/2$ when $\Delta\nu=0$. The first term in the
square bracket of Eq.~(\ref{Enuy2}) describes coherent spontaneous
radiation from a density-modulated beam and grows linearly with the
undulator distance $z_2$ (as discussed in Ref.~\cite{Yu} in the
context of harmonic generation). Since the electron beam from the
first undulator possesses not only density modulation but also
energy modulation, the momentum compaction of the second undulator
can convert the energy modulation into additional density
modulation. Thus, the second term in the square bracket of
Eq.~(\ref{Enuy2}) describes the enhanced radiation due to the
evolution of the density modulations inside the second undulator
which grows quadratically with the undulator distance.

In order to generate circularly polarized light, we require that
both $E_x$ and $E_y$ have the same average amplitude. From
Eq.~(\ref{Enuy2}), this corresponds to the condition
    \be\label{equalintnesity}
    \left\vert\frac{2i}{\mu_0^2}\left[\rho k_u z_2-\mu_0 e^{i\alpha}(\rho k_u z_2)^2
    \right]
    \right\vert=1\,.
    \ee
We consider a cold electron beam with vanishing energy spread, hence
the growth rate Im($\mu_0$) is maximized on resonance, i.e.,
$\Delta\nu=0$. In this case we obtain the required length of the
second undulator from Eq.(\ref{equalintnesity})
    \be\label{1p3Lg} L_2\approx1.3L_G\,, \quad \text{where} \quad
    L_G=\frac{\lambda_u}{4\pi\sqrt3\rho}
    \ee
is the 1D power gain length.

\section{Degree of Polarization}

The interference of the two radiation components generated by the
crossed undulator will produce flexible polarization. At the end of
the second undulator when $z=L_1+L_2$, these radiation fields in the
time domain are
    \begin{align} \label{eq:field t}
    E_y(t)=&\int {\omega_1 d\nu\over \sqrt{2\pi}} E_\nu^y (z_2=L_2) e^{i\Delta \nu \left[(k_1+k_u)
    (L_1+L_2)-\omega_1 t\right]}\,, \nonumber  \\
    E_x(t)=&\int {\omega_1 d\nu\over \sqrt{2\pi}} E_\nu^x (z=L_1) e^{i\Delta \nu \left[(k_1+k_u)
    L_1+k_1 L_2-\omega_1 t\right]}\,.
    \end{align}
Note that we only used Eq.~(\ref{eq:fourier}) for $E_x$ at $z=L_1$
(and $t_1$) and applied the free space propagation phase factor
$e^{i\Delta \nu \left[k_1 L_2-\omega_1 (t-t_1)\right]}$ in the
second undulator as $E_x$ does not interact with the electron beam
there. Because of the chaotic nature of SASE radiation, we perform a
statistical analysis to quantify the state of polarization.

Following the standard optics textbooks (see, e.g.,
Refs.~\cite{Born,Goodman}), the state of polarization can be
described by the coherency matrix
    \be\label{matrix1}
    \Jb =\left[
    \begin{array}{cc}
      \langle E_x(t) E_x^*(t)\rangle & \langle E_x(t) E_y^*(t)\rangle \\
      \langle E_y(t) E_x^*(t)\rangle & \langle E_y(t) E_y^*(t)\rangle\\
    \end{array}%
    \right]\,,
    \ee
where * means complex conjugate, and the angular bracket refers to
the ensemble average. The degree of polarization can be calculated
as~\cite{Born,Goodman}
    \be\label{degree of polarization1} P
    \equiv\sqrt{1-4\frac{\textrm{det}[\Jb]}{(\textrm{tr}[\Jb])^2}}\,,
    \ee
where $\textrm{det}[\Jb]$ and $\textrm{tr}[\Jb]$ are the determinant
and trace of the coherency matrix, respectively. It is also
convenient to introduce the first-order time correlation between
$E_x$ and $E_y$ as
\begin{align}\label{correlation_def}
    g_{xy}\equiv {\langle E_x(t) E_y^*(t)\rangle
    \over [\langle \vert E_x(t)\vert^2\rangle \langle \vert
    E_y(t)\vert^2\rangle]^{1/2}}.
    \end{align}

For polarization control in the crossed undulator, we are
particularly interested in the case when the average intensities of
the two radiation components are the same: $\langle \vert
E_x(t)\vert^2\rangle= \langle \vert E_y(t)\vert^2\rangle=\bar{I}$.
Under this condition, the coherency matrix simplifies to
    \be\label{matrix2}
    \Jb =\bar{I}\left[
    \begin{array}{cc}
                   1 & \vert g_{xy}\vert e^{i\theta} \\
       \vert g_{xy}\vert e^{-i\theta} & 1\\
    \end{array}%
    \right]\,,
    \ee
where $\theta$ is the phase difference between $E_x$ and $E_y$. When
$\theta=\pm\frac{\pi}{2}$, the combined radiation is circularly
polarized; when $\theta=0$ or $\pi$, it is linearly polarized at
$\pm45^\circ$ relative to the horizontal axis. The state of
polarization is controllable by adjusting the phase shift $\phi$ in
Eq.~(\ref{Enuy2}) so that the net phase in $g_{xy}$ is
$\theta=\pm\frac{\pi}{2}$ or $0/\pi$. With equal intensity in both
transverse directions, the degree of polarization in
Eq.~(\ref{degree of polarization1}) is simply given by the amplitude
of the $x$-$y$ time correlation, i.e.,
    \be\label{polarization_circ}
    P=\vert g_{xy}\vert\,.
    \ee

In the x-ray wavelength region, the electron bunch duration is
typically much longer than the coherence time of the SASE radiation.
Thus, a SASE pulse consists of many random intensity spikes that are
statistically independent. For a flattop current distribution (of
width $T$), we can convert the ensemble average of
Eq.~(\ref{correlation_def}) into a time average as
    \begin{align}
    g_{xy} =& \lim_{T\rightarrow\infty} {1\over \bar{I} T}
    \int_{-T/2}^{T/2} dt E_x(t) E_y^*(t)  \nonumber \\
    =& {1\over \bar{I} T}
    \int_{-\infty}^{\infty} \omega_1 d\nu E_\nu^x (L_1) E_\nu^{y*} (L_2) e^{-i\Delta\nu k_uL_2}
    \,,
    \end{align}
where we have applied Eq.~(\ref{eq:field t}) and the Parseval
relation in converting the time integration to the frequency
integration. Assuming that the first undulator operates in the
exponential gain regime, the frequency dependence of $E_\nu^x$ is
approximately Gaussian, i.e.,
    \be
    \langle\vert
    E_{\nu}^x(z)\vert^2\rangle=\frac{\bar{I}T}{\sqrt{2\pi}\sigma_\omega}e^{-\frac{(\Delta\nu)^2}{2\sigma_\nu^2}}\,,
    \ee
where the relative rms SASE bandwidth is~\cite{Kim_nim86,wang}
    \be \label{sase bw}
    \sigma_\nu=\sigma_\omega/\omega_1= \sqrt{9 \rho\over \sqrt{3} k_u
    L_1}\,.
    \ee
Since the short second undulator generates coherent radiation from a
pre-bunched beam that possesses the same narrow bandwidth
$\sigma_\nu$, we can expand $\mu_0^2$ in Eq.~(\ref{Enuy2}) to first
order in $\Delta \nu$ by using Eq.~(\ref{eq:mu0}). We also ignore
the frequency dependence of the second term in the square bracket of
Eq.~(\ref{Enuy2}) because its contribution to the radiation
intensity is relatively small. Finally, we have
    \begin{align}\label{correlation_result}
    \vert g_{xy}\vert \approx \frac{1}{\sqrt{2\pi}}\Bigg\vert \int_{-\infty}^\infty
    d\bar{\nu}\frac{\exp\left(-{\bar{\nu}^2\over 2}-i{\bar{\nu}\sigma_\nu k_uL_2\over 2}\right)
    \textrm{sinc}\left({\bar{\nu}\sigma_\nu k_uL_2\over 2}\right)}
    {1+(-\frac{1}{2}+i\frac{\sqrt{3}}{2}){\bar{\nu}\sigma_\nu\over 3\rho}}\Bigg\vert\,,
    \end{align}
where $\bar{\nu}=\Delta\nu/\sigma_\nu$. In view of
Eq.~(\ref{1p3Lg}), we take $L_2=1.3 L_G$ in
Eq.~(\ref{correlation_result}) and obtain the degree of polarization
by computing $\vert g_{xy}\vert$.

\section{Numerical simulations}
\label{sec:simu}

\subsection{1D results}

We first use a 1D FEL code to simulate the SASE radiation produced
by the crossed undulator configuration and to analyze the degree of
polarization. The code follows the time-dependent approach developed
in Ref.~\cite{bonifacio2} and employs the shot noise algorithm of
Penman and McNeil~\cite{penman}. Electron energy spread can be
included using Fawley's beamlet method~\cite{fawley}. After
computing the $E_x$ field produced in the first undulator, we allow
$E_x$ to propagate freely without further interacting with the
electron beam. The simulated electron distribution from the first
undulator is then used to generate the $E_y$ field in the second
undulator.

\begin{table}
\begin{center}
\caption{\label{tab:lcls}Main parameters for the LCLS soft x-ray FEL
used in simulations.}
\begin{tabular}{|l|c|c|}
\hline {\bf Parameter} & {\bf value} & {\bf unit} \\
\hline electron beam energy & 4.3 & GeV
\\ relative energy spread & 0(0.023) & \%
\\ bunch peak current & 2 & kA
\\ transverse norm. emittance & 1.2 & $\mu$m
\\ average beta function & 8 & m
\\ undulator period $\lambda_u$ & 3 & cm
\\ undulator parameter $K$ & 3.5 &
\\ FEL wavelength & 1.509 & nm
\\ FEL $\rho$ parameter & 0.119&  \%
\\ 1D power gain length $L_G$ & 1.17 & m
\\ 3D power gain length $L_G^{3D}$ & 1.48 & m
\\ \hline
\end{tabular}
\end{center}
\end{table}

As a numerical example, we use the parameter set listed in
Table~\ref{tab:lcls} that is similar to the soft x-ray LCLS
operation~\cite{LCLS}. In the 1D simulations, the energy spread is
set to zero since we want to compare with the previous analytical
results.

\begin{figure}[tb]
\centering
\includegraphics[scale=0.5]{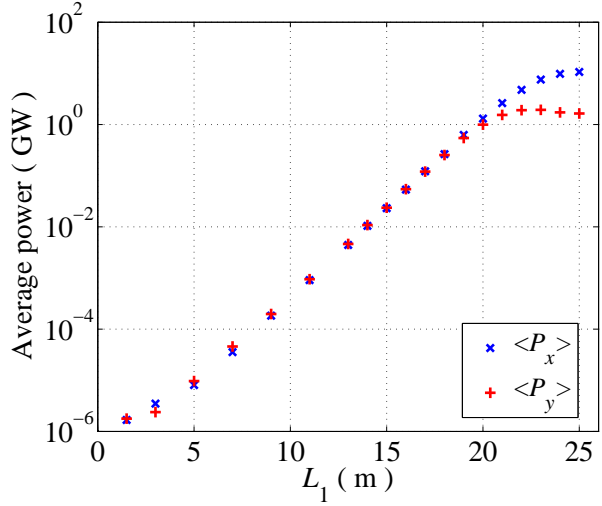}
\caption{(color) 1D simulations of the average SASE power at 1.5 nm
from the first (blue cross) and the second (red plus) undulator.
Here $L_1$ is the length of the first undulator, $L_2=1.3 L_G=1.53$
m is the length of the second undulator.} \label{xy_power}
\end{figure}

\begin{figure}[tb]
\centering
\includegraphics[scale=0.48]{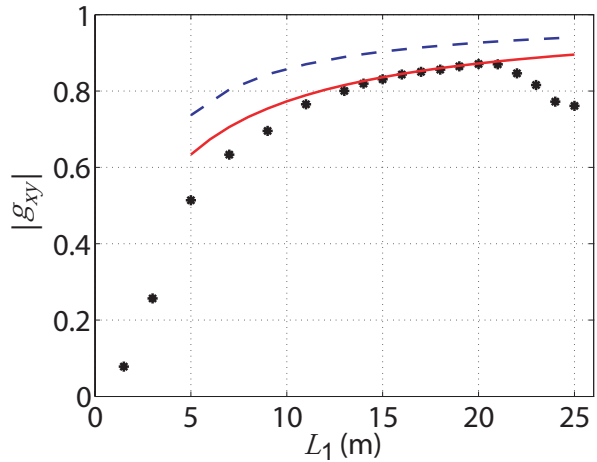}
\caption{(color) Amplitude of the time correlation $\vert
g_{xy}\vert$ from the 1D simulations (black star) and from
Eq.~(\ref{correlation_result}) (red solid curve). The degree of
polarization is equal to  $\vert g_{xy}\vert$ when $\langle P_x
\rangle=\langle P_x \rangle$. For comparison, the blue dashed curve
shows an estimate of the polarization by
Eq.~(\ref{correlation_simple}) (see text for more details). }
\label{correlation_fig}
\end{figure}

Fig.~\ref{xy_power} shows the average radiation power in both $x$
and $y$ directions produced by the cross undulator. The length of
the first undulator is allowed to vary, while the second undulator
length $L_2=1.3 L_G\approx 1.53$ m is held constant. As predicted by
Eq.~(\ref{1p3Lg}), the power of the two radiation components are
essentially the same in the exponential gain regime. Near
saturation, the power of the vertical field is lower than that of
the horizontal one because the FEL-induced energy spread starts to
de-bunch the electron beam in the second undulator. We repeat the
simulations 200 times for each $L_1$ with different random seeds to
start the process and calculate the first-order time correlation
between $E_x$ and $E_y$ at the exit of the second undulator using
the ensemble average defined in Eq.~(\ref{correlation_def}).
Figure~\ref{correlation_fig} shows the amplitude of this correlation
from the simulation results as well as the numerical integration of
Eq.~(\ref{correlation_result}) (the red solid curve) for a
comparison. When the first undulator is less than a couple of gain
lengths, the crossed undulator operates in the spontaneous emission
regime, the amplitude of the $x$-$y$ correlation and hence the
degree of polarization are very small without the use of a
monochromator. The degree of polarization increases in the
exponential growth regime and reaches a maximum of 85\% near the FEL
saturation. In this regime and especially when the gain is very
high, we see very good agreement between simulations and
Eq.~(\ref{correlation_result}). In the saturation regime, the
amplitude of the $x$-$y$ correlation starts to decrease, and the
linear theory starts to deviate from the simulation results.

There are two effects that prevent the degree of polarization to
reach 100\% in a crossed-undulator SASE FEL. First, there is
relative slippage between $E_x$ and $E_y$ in the second undulator.
Since $E_x$ stops interacting with the electron beam after the first
undulator, the group velocity of $E_x$ is the speed of light $c$.
However, the group velocity of $E_y$ is slower than $c$ because it
is generated by the micro-bunched beam that travels at the average
longitudinal velocity $\beta_\parallel c$. In fact, 1D simulations
indicate that the group velocity of $E_y$ is almost the same as that
of the electrons within the short second undulator section. (This
numerical result is also confirmed in 3D simulations to be discussed
in the next section.) To estimate the slippage effect, we take
$E_y(t)\approx E_x(t-\tau)$ with $c\tau=L_2(1-\beta_\parallel)$ and
apply the first-order time correlation function of the SASE field to
estimate $\vert g_{xy}\vert$:
   \begin{align}\label{correlation_simple}
   g(\tau) =\textrm{exp}\left(-{\pi\tau^2\over 2{\tau_c}^2}\right)\,,
   \end{align}
where $\tau_c=\sqrt{\pi}/{\sigma_\omega}$ is the coherence
time~\cite{saldin}, and $\sigma_\omega$ is given by Eq.~(\ref{sase
bw}). Equation~(\ref{correlation_simple}) yields the blue dashed
curve shown in Fig.~\ref{correlation_fig}, which indicates that the
slippage effect only accounts for about a half of the depolarization
in the crossed undulator. A careful examination of the intensity
profile between $E_x$ and $E_y$ shows a visible difference from a
simple time delay (see Fig.~\ref{3Dprofile} for a 3D example). This
accounts for the additional depolarization effect in a crossed
undulator SASE FEL.

\subsection{3D Discussions}

A remarkable feature of a SASE FEL is its transverse coherence. At a
sufficiently high gain, a single transverse mode with the largest
growth rate will dominate over all other transverse modes for a
typical SASE FEL. Thus, we expect the previous 1D analysis still
applies to 3D situations in the high gain limit, with the maximum
polarization obtainable at the end of the exponential growth regime.
Since the length of the second undulator is short, the diffraction
effects for the free-propagating $E_x$ in the x-ray wavelength
regime is expected to be small. Thus, the 3D effects such as
emittance and diffraction do not play significant roles in
determining the degree of polarization for a crossed undulator SASE
FEL.

\begin{figure}[tb]
\centering
\includegraphics[scale=0.5]{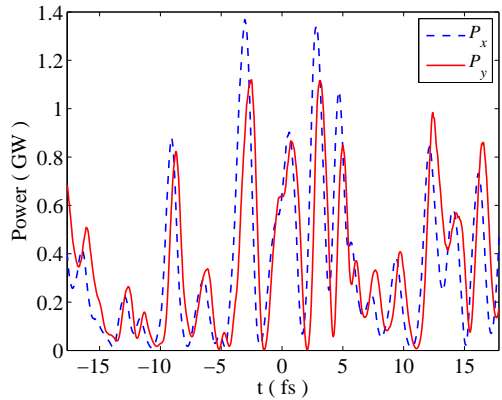}
\caption{(color) GENESIS simulated power profiles of the horizontal
field (blue dashed curve) and the vertical field (red solid curve)
at the end of the second undulator, bunch head at left. }
\label{3Dprofile}
\end{figure}

We use the 3D FEL code GENESIS 1.3~\cite{genesis} to check these
expectations. The electron beam is dumped at the end of the first
undulator and is used to generate $E_y$ in the second undulator.
$E_x$ propagates in the same length of the second undulator but
without any undulator magnetic field. We use the same soft x-ray FEL
example listed in Table~\ref{tab:lcls} as the 1D case but with a
relative energy spread of 0.023\%, which roughly corresponds to the
LCLS soft x-ray parameters. The length of the first undulator is
chosen to be 23 m and is about 3 m before the saturation point. A
2-m short second undulator is necessary to produce the same
radiation power for the vertical field (see Fig.~\ref{3Dprofile}).
The 3D power gain length corresponding to these parameters is
$L_G^{3D}=1.48$~m, so Eq.~(\ref{1p3Lg}) approximately holds in this
3D case. We use the on-axis far-field radiation intensity and phase
from GENESIS simulations to calculate the time correlation between
$E_x$ and $E_y$ of Eq.~(\ref{correlation_def}). Instead of
performing many statistical runs for the ensemble average, we
average the result over hundreds of intensity spikes within the
radiation pulse in order to save on simulation effort. The amplitude
of the $x$-$y$ correlation from this 3D calculation is 87\%, very
close to the 1D prediction. Figure~\ref{3Dprofile} shows the central
section of the simulated power profiles $P_x$ and $P_y$ at the end
of the second undulator. A small time delay due to the slippage
effect and a somewhat different temporal structures between $P_x$
and $P_y$ are the main depolarization effects, as discussed in the
previous section.

\section{Conclusions}

The statistical analysis presented in this paper shows that the
crossed-planar undulator is an effective method for polarization
control in a SASE FEL. To optimize the degree of polarization, the
first undulator should operate at the end of the exponential growth
regime, while in order to generate circularly polarized x-rays, the
second undulator should be about 1.3 times the power gain length.
The maximum degree of polarization is over 80\% from both theory and
simulations. If fast pulsed magnets are employed in the phase
shifter chicane, the relative phase between the two radiation
components from the crossed undulator can vary at hundreds of Hz,
hence enabling fast polarization switching for many scientific
applications.

\section{acknowledgments}

We thank P. Emma, J. Hastings, and K.-J. Kim for many useful
discussions. Special thanks to K. Bane for a careful reading of the
manuscript and for his comments. This work was supported by
Department of Energy Contract No. DE-AC02-76SF00515.

\end{document}